\documentclass[12pt]{article}

\usepackage{newtxtext,newtxmath}

\usepackage{graphicx}
\usepackage{xcolor}
\usepackage{xspace}

\usepackage[letterpaper,margin=1in]{geometry}

\usepackage{hyperref}

\renewenvironment{abstract}
	{\quotation}
	{\endquotation}

\date{}

\makeatletter
\renewcommand{\fnum@figure}{\textbf{Figure \thefigure}}
\renewcommand{\fnum@table}{\textbf{Table \thetable}}
\makeatother

\usepackage{scicite}

\usepackage{url}
\usepackage{authblk}
\newcommand{\ceco}{CeCoIn$_5$\xspace}
\usepackage{xfrac}

\usepackage{braket}
\usepackage{bm}

\def\scititle{
	Neutron Spin Resonance as Probe of Triplet Pair Density Waves in CeCoIn$_5$
}
\title{\bfseries \boldmath \scititle}

\author{
    David~W.~Tam$^{1,2,3\ast\dagger}$,
    Ilaria~Maccari$^{4,5\dagger}$,
    St\'ephane~Raymond$^{6,\ast}$,
    Barry~Winn$^{7}$,\\
    Paul~Steffens$^{1}$,
    Arno~Hiess$^{1,8}$,
    Jacques~Ollivier$^{1}$,
    Dariusz~Jakub~Gawryluk$^{2}$,\\
    Ekaterina~Pomjakushina$^{2}$,
    Aline~Ramires$^{5,9,\ast}$,
    Michel~Kenzelmann$^{2,\ast}$\and
\small$^{1}$Institut Laue-Langevin, 71 avenue des Martyrs, 38042 Grenoble Cedex 9, France. \\
\small$^{2}$PSI Center for Neutron and Muon Sciences, 5232 Villigen PSI, Switzerland.\\
\small$^{3}$Department of Applied Physics, KTH Royal Institute of Technology, SE-106 91 Stockholm, Sweden. \\
\small$^{4}$Institute for Theoretical Physics, ETH Zurich, 8093 Zurich, Switzerland.\\
\small$^{5}$PSI Center for Scientific Computing, Theory, and Data, Paul Scherrer Institute, 5232 Villigen PSI, Switzerland.\\
\small$^{6}$Université Grenoble Alpes, CEA, IRIG, MEM, MDN, 38000 Grenoble, France.\\
\small$^{7}$Neutron Scattering Division, Oak Ridge National Laboratory, Oak Ridge, TN 37831, USA.\\
\small$^{8}$European Spallation Source ERIC, P.O. Box 176, 221 00 Lund, Sweden.\\
\small$^{^9}$Institute of Solid State Physics, TU Wien, 1040 Vienna, Austria.\\
\small$^\ast$Corresponding authors. Emails: tamd@ill.fr, raymond@ill.fr,imaccari@phys.ethz.ch, aline.ramires@tuwien.ac.at, michel.kenzelmann@psi.ch\\
	\small$^\dagger$These authors contributed equally to this work.
}

\begin{document} 

\maketitle

\begin{abstract} \bfseries \boldmath



Spin resonance modes are among the most characteristic collective excitations of unconventional superconductors, yet their microscopic origin remains unresolved. 
Although commonly interpreted as purely magnetic excitations, this picture does not explain the rich magnetic field evolution of these modes observed in heavy-fermion superconductors. 
Here we combine inelastic neutron scattering and symmetry analysis to investigate the evolution of the spin resonance modes in the heavy-fermion superconductor \ceco under an in-plane magnetic field. 
At high fields, approaching the upper critical field of the superconductor, we discover a third resonance mode with relatively high intensity.
We demonstrate that the resonance branches observed experimentally can be interpreted as collective modes of the superconducting condensate involving transitions to excited superconducting states.
This framework provides a unified interpretation of the field-induced splitting of the resonance, the newly observed mode at high fields, the relative neutron scattering intensities of each mode, and provides the first direct evidence that a spin resonance mode can be associated with a singlet-triplet excitation of a superconductor.
\end{abstract}


Spin resonance modes (SRMs) are among the most prominent collective excitations of unconventional superconductors~\cite{Tranquada2014}. 
Observed only below the superconducting transition temperature in cuprates~\cite{Sidis2004, Fujita2012}, iron-based superconductors~\cite{Lumsden2010, Dai2015}, and heavy-fermion compounds~\cite{Stockert2023}, they provide a unique window into the interplay between magnetism and superconductivity. 
Their energies exhibit suggestive scaling relations with the superconducting gap, raising the possibility that they encode universal information about the pairing mechanism~\cite{Yu2009}. 
Yet, despite decades of experimental and theoretical effort, a unified microscopic understanding of spin resonance modes across different classes of unconventional superconductors remains elusive.

\begin{figure*}[h]
\includegraphics[width=.95\textwidth]{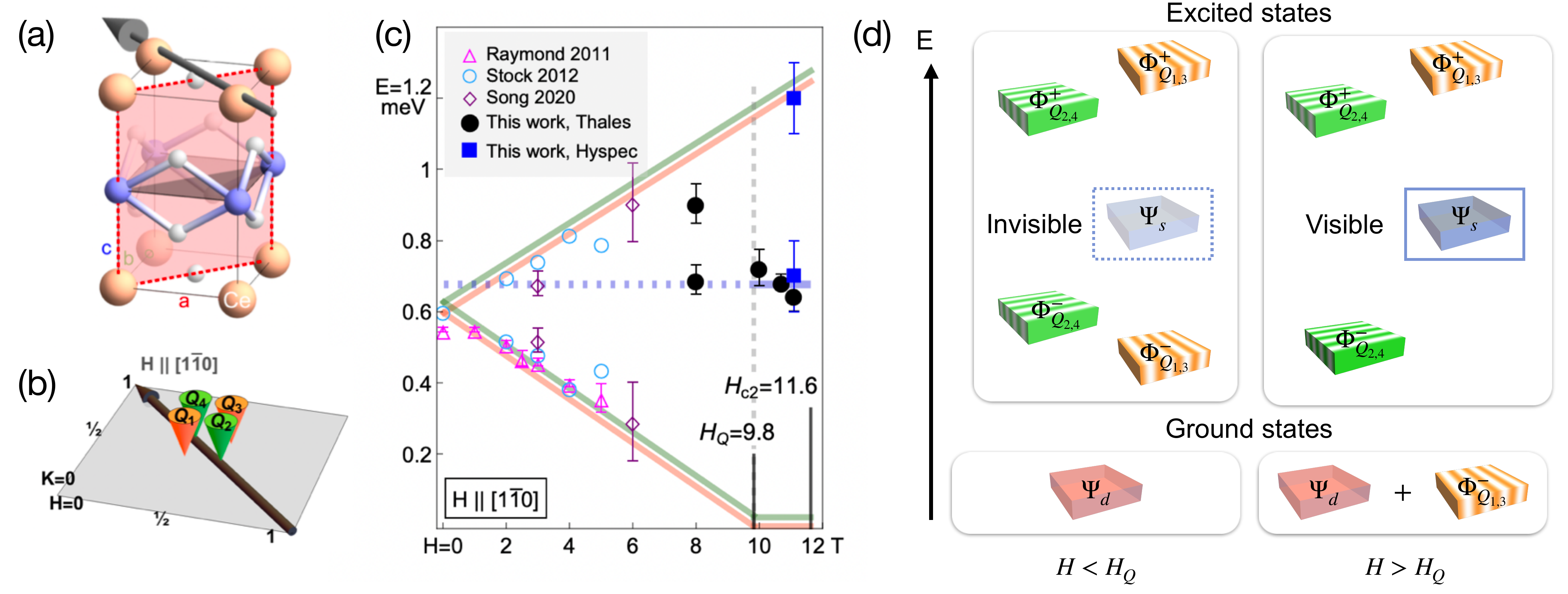}
\caption{ 
\textbf{ A new spin resonance mode at high magnetic fields.}
\textbf{(a)} Crystal structure of \ceco highlighting the mirror plane that remains as a symmetry in the presence of a magnetic field along the $[1\bar{1}0]$ direction (grey arrow). 
\textbf{(b)} Schematic of the wave vectors identified in the inelastic neutron scattering (INS) experiments. The conical shape represents the upward dispersion stemming from each of the $Q_i$.
\textbf{(c)} Spin resonance modes observed by multiple INS experiments \cite{raymondEvolutionSpinResonance2011,Stock2012,Song2020}, and highlighting the new branch for $H>H_Q$.
\textbf{(d)} Schematic of the possible excitations in the superconducting state. For $H<H_Q$, the superconducting ground state consists of only a d-wave spin-singlet component ($\Psi_d$), and excitations can be observed to all modulated spin-triplet excited states ($\Phi^{\pm}_{Q_i}$) but not to the homogeneous s-wave singlet state ($\Psi_s$). For $H>H_Q$, the ground state includes a modulated component with spin-triplet character ($\Phi^{-}_{Q_{1,3}}$), which makes the central mode corresponding to the homogeneous s-wave singlet state visible.
}
\label{fig1}
\end{figure*}

The tetragonal heavy-fermion superconductor \ceco (Fig.~\ref{fig1} (a)) provides a particularly compelling platform to address this unresolved issue. 
It is a clean, quasi-two-dimensional unconventional superconductor with $T_c \simeq 2.3$ K and a well-established $d_{x^2-y^2}$-wave order parameter \cite{Petrovic2001, Settai2001, Movshovich2002, Kasahara2005, allanImagingCooperPairing2013}. 
Unlike many unconventional superconductors, its upper critical field is Pauli limited, allowing the spin resonance to be studied in a wide range of magnetic fields while remaining within the superconducting state \cite{Bianchi2002,kenzelmannExoticMagneticStates2017}. 
At zero field, CeCoIn$_5$ exhibits no static magnetic order, providing a simple setting in which to investigate the coupling between superconductivity and magnetic excitations.

The SRM in CeCoIn$_5$ emerges below $T_c$ at an energy of approximately $0.6$ meV near the wavevector $\mathbf{Q}=(1/2,1/2,1/2)$~\cite{Stock2008}.
Under an in-plane magnetic field, the SRM splits into distinct branches, with the lowest-energy branch decreasing approximately linearly with field and softening towards zero energy at the onset of the Q phase~\cite{Stock2012}. 
The Q phase has been characterized as a field-induced incommensurate spin density wave (SDW) state that coexists with superconductivity up to its upper critical field~\cite{Bianchi2002,
Kenzelmann2008, Kenzelmann2010}.
While these observations have motivated an interpretation of the SRM as a precursor of the field-induced SDW order~\cite{Raymond2015}, the microscopic origin of the SRM remains under debate. 
In particular, neutron-scattering measurements have challenged the conventional spin-exciton interpretation. The observed upward dispersion, rather than the expected downward-dispersing branch, and the partial suppression of low-energy magnetic fluctuations below $T_c$ have instead motivated an interpretation in terms of a pre-existing, magnon-like magnetic excitation whose damping is reduced in the superconducting state~\cite{Song2016, Song2020}. These different scenarios highlight the unresolved relationship among superconducting degrees of freedom, the magnetic instability, and the collective excitations in CeCoIn$_5$.
One tantalizing possibility is that its collective character can instead be understood directly in terms of the superconducting order parameter and its excitations, since it has been recognized that high field states may tend to favor triplet pairing in the general case \cite{shimahara_coexistence_2000}.


\textbf{Spin resonance splitting beyond wavevector degeneracy.} 
We first examine its evolution under an in-plane magnetic field at low fields.
At zero field, neutron scattering reveals a resonance at approximately $0.6$~meV, distributed over four symmetry-related satellites at wavevectors $\mathbf{Q}_{1-4} = (1/2\pm \delta,1/2\pm\delta,1/2)$, with $\delta \approx 0.06$, as sketched in Fig. 1~(b) \cite{Raymond2015}. 
Upon applying an in-plane magnetic field, the resonance splits into two branches (Fig. 1~(c)) \cite{Stock2012, Song2020}. Inelastic neutron scattering with $H \| [1\bar{1}0]$ at $H=4$~T (Fig.~2) clearly shows that all four satellites are present in the lower branch near $E=0.35$~meV.
This demonstrates the important result that the magnetic-field splitting of the SRM into two branches cannot be attributed to the lifting of the degeneracy between the different ordering wavevectors, parallel or perpendicular to the applied field.
Instead, the field reveals an additional internal degree of freedom of the resonance beyond its magnetic ordering wavevector.


\textbf{A distinct resonance emerges inside the Q phase.} 
As the magnetic field is further increased, the lower-energy branch softens to zero energy at $H_Q=9.8$~T, coinciding with the onset of the Q phase. 
Above $H_Q$, however, we find evidence of higher-energy modes.
Most significantly, we observe the emergence of a new branch whose energy does not shift in the field and is close to the energy of the SRM in zero field.
We refer to this as the ``central'' branch, as its position is lower than that of an upward-shifting branch that we identify in both low- and high-field measurements \cite{Stock2012, Song2020}.
In Fig.~1(c), we diagram the position of the central mode obtained from two experiments with the time-of-flight and triple-axis techniques (see Methods).
From the time-of-flight experiments, we show that the momentum-space structure of the central mode is centered near $\mathbf{Q}$, and it disappears above $H_{c2}$ at $H=12$~T, clearly showing its association with the superconducting state (Fig. 3). 
Moreover, energy-momentum cuts reveal a distinctive upward-dispersive motif moving along the $(H,H)$ direction, parallel to $[110]$.
Since a similar upward-dispersing feature was observed in previous studies at lower fields \cite{Song2020}, the appearance of the same characteristics provides further evidence that this new central branch is a spin resonance mode branch.


\textbf{SRMs as collective excitations of the superconducting state.}
To account for these observations, we develop a theoretical framework in which the SRMs correspond to collective excitations of the superconducting condensate, rather than a distinct magnetic instability. 
We interpret the resonance branches as transitions from the homogeneous spin-singlet superconducting ground state to subdominant pairing channels, in the spirit of the Bardasis–Schrieffer picture \cite{Bardasis1961}. 
Within this framework, an applied magnetic field lifts degeneracies and continuously reshapes the spectrum of collective excitations. 
Neutron-scattering selection rules then determine which excitations are observable. 
Within our framework, the relevant scattering intensity is governed by matrix elements of the magnetization density operator, $ \bra{\lambda} \hat{\mathbf{S}}(\mathbf{Q}) \ket{\lambda'}$, between the superconducting ground state, $\ket{\lambda}$, and excited state, $\ket{\lambda'}$  (see Methods and Supplementary information \cite{SI} for details). 
Since the superconducting ground state at low magnetic fields is a spin singlet, a nonzero matrix element requires the excited state to carry a spin-triplet component.
Thus, neutron scattering is directly sensitive to the singlet or triplet character of the excited state branches.


\textbf{Properties of the superconducting ground states}.
To identify the excited pairing channels responsible for the observed SRMs, we first note that \ceco has tetragonal symmetry and realizes a homogeneous $B_{1g}$ superconducting state at zero field, with form factor $\propto k_x^2-k_y^2$, which we denote as $\Psi_{B_{1g}}$.
Building on proposals that associate the magnetically ordered Q-phase with a pair density wave (PDW) order~\cite{Aperis2008, Miyake2008, Agterberg2009, Aperis2010}, we assume that the SDW is induced by the coexistence of the homogeneous $B_{1g}$ superconducting state and an underlying PDW. 

This conjecture allows the symmetry of the PDW contributing to the Q-phase to be uniquely determined.
For concreteness, we focus on the PDW with modulating wavevectors $\mathbf{Q}_{1,3}$, which are perpendicular to the field (Fig.~1(b)).
Once the magnetic field $H_{\bar{d}}$ is applied along the antidiagonal direction $\bar{d} \equiv [1\bar{1}0]$, the only symmetry preserving the wavevector, and therefore useful for classifying the internal structure of the order parameter, is the mirror $M_{\bar{d}}$ with normal along $\bar{d}$ (Fig.~1 (a)).
We denote the corresponding PDW order parameters by $\Psi_{\mathbf{Q}_{1,3}}^{e \backslash o}$, according to their even ($e$) or odd ($o$) character under $M_{\bar{d}}$, and the SDW order parameter as $S_{\mathbf{Q}_{1,3}}^z$.
The free energy term responsible for the induction of the SDW has the form $ \propto iS_{+\mathbf{Q}_{1,3}}^z \left( \Psi_{B_{1g}}^* \Psi_{-\mathbf{Q}_{1,3}}^{e} - \Psi_{B_{1g}} \Psi_{+\mathbf{Q}_{1,3}}^{e  *} \right) + c.c.$ \cite{Agterberg2009}. 
Because both $\Psi_{B_{1g}}$ and the observed SDW are $M_{\bar{d}}$-odd, symmetry requires the PDW component entering this coupling to be $M_{\bar{d}}$-even.
Thus, the symmetry of the PDW contributing to the Q phase is uniquely determined. {The same conclusion follows from a second symmetry-allowed term, $\propto H_{\bar{d}}\, S^z_{+\mathbf{Q}_{1,3}} \left( \Psi^*_{B_{1g}} \Psi^{e}_{-\mathbf{Q}_{1,3}} + \Psi_{B_{1g}} \Psi^{e*}_{+\mathbf{Q}_{1,3}} \right) + \mathrm{c.c.}$, which is linear in the magnetic field. 
Since $H_{\bar{d}}$ is itself $M_{\bar{d}}$-even, it leaves the mirror parity of the coupling unchanged, so the PDW component must again be $M_{\bar{d}}$-even.}

This symmetry analysis also allows us to understand why, while the SRMs display intensities in all wavevectors $\mathbf{Q}_{1,..,4}$, only the modes with $\mathbf{Q}_{1,3}$ condense at $H_Q$ when the magnetic field is applied along the $[1\bar{1}0]$ direction \cite{Gerber2013}.
The applied magnetic field breaks the equivalence between the diagonal and anti-diagonal directions and, consequently, between modulated states with wave vectors parallel, $\mathbf{Q}_{2,4}$, and perpendicular, $\mathbf{Q}_{1.3}$, to it.
This distinction and wavevector selectivity can be traced to the symmetry of the underlying homogeneous superconducting state. 
In zero field, the two diagonal mirror planes, $M_d$ and $M_{\bar{d}}$, protect the nodes of the superconducting order parameter along their respective planes. 
A field along $[1\bar{1}0]$ breaks the $M_d$ mirror symmetry and therefore lifts the symmetry protection of the nodes along the anti-diagonal direction, while the nodes along the diagonal remain protected by $M_{\bar d}$ (Fig.~4 (a)). 
The resulting low-energy spectrum is then gapped along $\mathbf{Q}_{2,4}$, but retains low-lying quasiparticles along $\mathbf{Q}_{1,3}$, favouring modulated orders with wave vectors along $\mathbf{Q}_{1,3}$. 
Although the resulting energy difference between the two sets of wavevectors is too small to be resolved in the inelastic spectra, it can select the ordering wavevector at the phase transition.

\textbf{Excited PDW states at low fields}.
While PDWs generally incorporate both spin-singlet and spin-triplet contributions, since the neutron-scattering selection rules require a spin-triplet excited state for nonzero intensity, we now focus on the latter.
In the presence of a magnetic field along the $[1\bar{1}0]$ direction, the spin-triplet content of the superconducting order parameter can acquire components in all three directions $\mathbf{\Psi}^e_{\mathbf{Q}_{1,3}}(\mathbf{k}) = [\Psi_d^e(\mathbf{k}), \Psi_{\bar{d}}^e(\mathbf{k}), \Psi_z^e(\mathbf{k})]_{\mathbf{Q_{1,3}}}^T$ (see details in the SI \cite{SI}).
The multi-component nature of the spin-triplet superconducting order parameter allows for different order parameter configurations, some of which can be non-unitary~\cite{Ramires2022}, and leads to a general form of quadratic terms in the Landau free energy (omitting the $\mathbf{k}$ dependence) as $f = \sum_s f_{\mathbf{Q}_s}$ with  $f_{\mathbf Q_s}=\mathbf{\Psi}^{e\dagger}_{\mathbf{Q_{s}}}\mathcal M^{\mathbf{Q_{s}}}_H \mathbf{\Psi}^e_{\mathbf{Q_{s}}}$, where, for $\mathbf{Q}_{1,3}$, 
\begin{equation}
\mathcal M_{H}^{\mathbf{Q}_{1,3}}=\begin{pmatrix}
\alpha_d & \alpha_{d\bar{d}} & i\lambda \,H_{\bar d}\\[3pt]
\alpha_{d\bar{d}} & \alpha_{\bar{d}} + \gamma H_{\bar d}^{2} & 0\\[3pt]
-i\lambda\,H_{\bar d} & 0 & \alpha_z
\end{pmatrix}.
\label{eq:3x3_matrix}
\end{equation}
We use this matrix as an effective phenomenological description of the spin resonance modes whose eigenvalues track the excitation energies of the corresponding modes, with the aim of capturing their field evolution and mixing.
Here $\alpha_{i=\{d,\bar{d},z\}}>0$ encodes the relative energy with respect to the ground state for each PDW spin-triplet component in isolation, and $\alpha_{d\bar{d}}$ captures a symmetry-allowed hybridization between $\Psi_d$ and $\Psi_{\bar{d}}$ in zero field. 
The term $\gamma H_{\bar{d}}^2$, with $\gamma>0$, implements the Pauli limiting effect for the d-vector component parallel to the magnetic field, while the linear-in-field term $i\lambda H_{\bar{d}}$ describes the coupling of the magnetic field to the gauge-invariant superconducting bilinear for a nonunitary order parameter (see SI \cite{SI} for details).

Diagonalizing this matrix yields three orthogonal collective modes with field-dependent internal structures. 
Two of them correspond to the observed spin resonance modes: the lower branch, whose energy decreases with increasing field, and the upper branch, whose energy increases; we denote them $\Phi^-_{\mathbf{Q}_i}$ and $\Phi^+_{\mathbf{Q}_i}$, respectively.
Their eigenvectors determine the weights of the three d-vector components and, therefore, their contribution to the neutron scattering intensity, while the corresponding eigenvalues determine their field-dependent energies. 
We find that the field-induced splitting originates from the coupling between the in-plane and out-of-plane components of the triplet order parameter, as diagrammed in Fig.~4 (see SI \cite{SI} for more details).
Interestingly, our framework predicts a third mode at higher energy, which could be associated with an extra feature in the inelastic neutron scattering spectra between $1.5$ and $1.6$ meV that has never been previously identified,  see Fig.~3 (g).
Moreover, the predicted neutron scattering intensities in our model display an intrinsic asymmetry between the upper and lower branches stemming from the different weight of d-vector components, see Fig.~4 (b) and (c).
{For a representative choice of parameters, we can reproduce both the linear evolution of the energies of the two branches, $\Phi^-_{\mathbf{Q}_i}$ and $\Phi^+_{\mathbf{Q}_i}$, with magnetic field and their relative intensity, with $\Phi^-_{\mathbf{Q}_i}$ about two times more intense than $\Phi^+_{\mathbf{Q}_i}$, in agreement with previous experiments~\cite{Stock2012,Song2020}, see  Fig.~4 (d).}

\textbf{The emergence of a new central resonance mode.}
With regard to the new ``central'' mode observed in our high-field experiments (Fig.~3), the condensation of a PDW with spin-triplet correlations naturally opens a new neutron scattering channel, in which the condensed PDW is the initial state and the excited state has spin-singlet character (Fig.~1 (d)). 
In the presence of an external field along the $[1\bar{1}0]$ direction, the only other spin-singlet channel has $A_{g}$ symmetry.
We therefore associate the central branch with an excited homogeneous spin-singlet state in the $A_g$ representation. For simplicity, we consider a featureless s-wave form factor, which provides a nonzero neutron-scattering matrix element for transitions from the condensed triplet PDW.
The assignment of the central branch to a superconducting excited state with spin-singlet character can also account for its nearly field-independent energy, since both the ground and excited-state homogeneous orders are similarly suppressed by magnetic fields.
Evaluating the neutron scattering intensity corresponding to excitations from the spin-triplet components of the PDW to the homogeneous s-wave spin-singlet state, we find an almost constant intensity as a function of magnetic field, in agreement with our triple-axis neutron scattering data between $H=8-11.1$~T (Fig.~3). 
The appearance of the central branch below $H_Q$ in our experiments may be related to previous observations of strong magnetic fluctuations before the onset of the $Q$ phase \cite{koutroulakisFieldEvolutionCoexisting2010}, which is likely due to the thermal broadening of the transition at a finite temperature of $T=0.04$~K.

\textbf{Discussion and outlook.} 
Theoretical descriptions of the Q phase in CeCoIn$_5$ have explored several mechanisms involving the interplay between superconductivity and magnetism. 
These range from scenarios in which finite-momentum pairing coupled to homogeneous $d_{x^2-y^2}$-wave superconductivity induces SDW order~\cite{Miyake2008, Agterberg2009, Aperis2010}, to mechanisms in which Pauli-limited superconductivity, quasiparticle nesting, or vortex-lattice effects promote a magnetic instability~\cite{Kato2011, Michal2011}. 
However, the microscopic origin of the SRMs remains under debate. 

Our results suggest a new interpretation in which the SRMs are collective excitations of the superconducting condensate. 
In this framework, the resonance branches correspond to transitions between the superconducting ground state and subdominant pairing channels, extending the Bardasis–Schrieffer picture to the spin-sensitive response probed by neutron scattering. 
The field-induced splitting reflects the coupling between distinct internal components of the superconducting excitations, rather than a lifting of the degeneracy between different ordering wavevectors. Moreover, the neutron-scattering selection rules impose an independent constraint on the spin character of these excitations, requiring spin-triplet contributions to the excited states at low fields and to the ground state at high fields.

Our newly proposed approach to characterize spin resonance modes, based on collective excitations within the superconducting state, might provide a new way to identify their nature in other unconventional superconductors.
More broadly, our framework establishes that SRMs can be direct spectroscopic probes of hidden spin-triplet superconducting correlations in the excited state of unconventional superconductors, which has never been firmly established in any superconducting material up to now.
A microscopic theory will, however, be needed to establish the quantitative relationship between these proposed collective excitations and the measured resonance energies.

\section*{Methods}

\textbf{Experimental details.}
\ceco single crystals were grown using an indium flux method that we described previously \cite{Tam2023}.
Crystals were separated from the flux by centrifugation and later etched in hydrochloric acid.
The \ceco samples exhibited a plate-like habit corresponding to the basal plane of the crystal.
Samples were individually co-aligned using an x-ray Laue machine.

Inelastic neutron scattering experiments were carried out at $H=0-4$~T at IN5 (Institut Laue-Langevin, France) \cite{in5}, $H=11.1-12$~T at Hyspec (Spallation Neutron Source, Oak Ridge, USA), and $H=8-12$~T at Thales (Institut Laue-Langevin, France) \cite{thales}.
Our sample was an array of crystals of mass 0.7~g (IN5, Thales) or 1.4~g (Hyspec), which were co-aligned with a sample mosaic less than 4${}^\circ$ in all directions (see Supplementary information \cite{SI} for details).
In all experiments, the field was applied vertically along the $[1\bar{1}0]$ direction, in the basal plane of the sample.
This geometry ensures a fixed field direction in the frame of the sample, even as the sample was rotated in order to measure excitation throughout the horizontal $[110]\times[001]$ scattering plane.
This geometry ensures that the magnetic field remains in the basal plane of the crystal along $[1\bar{1}0]$ for all measurement configurations, even as the sample is rotated to collect data throughout the 
At IN5, the data was collected with $\lambda=4.8$~\AA~with Fermi speed 9600 rpm.
At Hyspec, data was collected with $E_i=5.1$~meV with Fermi speed 240 Hz.
The detector vessel was oriented to receive neutrons at $2\theta$ between -3 and -63 degrees.
At Thales, all data was collected with $k_f=1.5$~\AA.

\textbf{Inelastic Neutron Scattering matrix elements.}
Taking inspiration from the picture of SMR in systems of dimers~\cite{Lake1996,Lake1997} we treat the scattering states as pairs of electrons with given total orbital and spin angular momentum, expanded as
\begin{equation}
\ket{\lambda} = \sum_{L, M_L} \sum_{S, M_S}
\lambda^{LS}_{M_L M_S}\,
\ket{L, M_L} \otimes \ket{S, M_S},
\label{eq:state_expansion}
\end{equation}
where the coefficients $\lambda^{LS}_{M_L M_S}$ encode the orbital and
spin structure of each state. The transition matrix elements governing
the differential cross-section of inelastic neutron scattering then
read \cite{Lovesey1986}
\begin{eqnarray}
\bra{\lambda} \hat{\mathbf{S}}(\mathbf{Q}) \ket{\lambda'}
&=& \sum_{\substack{L, M_L, S, M_S \\ L', M'_L, S', M'_S}}
\left(\lambda^{LS}_{M_L M_S}\right)^{*}
\lambda'^{\,L'S'}_{M'_L M'_S}
\sum_{k, q} 4\pi\, i^{k}\,
Y_{k}^{q*}(\hat{\mathbf{Q}})\,
I_{3Y}(L, M_L; L', M'_L; k, q) \\
\nonumber&&\times {j}_k(Q)\,
\bra{S, M_S}\, \hat{\mathbf{S}}\, \ket{S', M'_S},
\label{eq:INS_matrix}
\end{eqnarray}
where $\hat{\mathbf{S}}$ is the spin operator, $I_{3Y} (L, M_L; L', M_L'; k,q)\equiv \int d\Omega_r \,  Y_L^{M_L *}(\hat{r})Y_{k}^q(\hat{r}) Y_{L'}^{M'_L}(\hat{r})$ enforces angular momentum selection rules, and ${j}_k(Q)$ is the radial overlap integral dependent on the magnitude of the wavevector $Q=|\mathbf{Q}|$. As we do not have a microscopic model to evaluate the radial integrals, their implicit values are used as units of the neutron scattering intensity. 

In the low-field regime, $H < H_Q$, the system scatters out of the homogeneous, spin-singlet $d_{x^2 - y^2}$ superconducting ground state, specified by
{
\begin{eqnarray}
  \ket{\lambda}_{H<H_Q} = \ket{\Psi_{B_{1g}}} = \ket{k_x^2 - k_y^2}\otimes\ket{s},
\end{eqnarray}
with the orbital (L=2) and spin (S=0) states:
\begin{eqnarray}
  \ket{k_x^2 - k_y^2} = \tfrac{1}{\sqrt2}\big[\ket{L{=}2,m{=}{-2}} + \ket{L{=}2,m{=}2}\big]; \quad
  \ket{s} = \ket{S{=}0,M_S{=}0} = \tfrac{1}{\sqrt2}\big(\ket{\uparrow\downarrow}-\ket{\downarrow\uparrow}\big).
\end{eqnarray}
}
%
The final state $\ket{\lambda'}$ corresponds to a superposition of the subdominant,
spatially modulated spin-triplet PDW configurations with $M_{\bar d}$-even symmetry:
{\begin{eqnarray}
  \ket{\Psi_d^e}       = \ket{k_x - k_y}\otimes\ket{\hat x + \hat y}; \quad
  \ket{\Psi_{\bar d}^e} = \ket{k_x + k_y}\otimes\ket{\hat x - \hat y}; \quad
  \ket{\Psi_z^e}       = \ket{k_x - k_y}\otimes\ket{\hat z};
\end{eqnarray}
where the orbital (L=1) and spin (S=1) states are
\begin{eqnarray}
  &&\ket{k_x \pm k_y} = \tfrac{1}{2}\big[(1\mp i)\ket{L{=}1,m{=}{-1}} - (1\pm i)\ket{L{=}1,m{=}1}\big],\\
  &&\ket{\hat x \pm \hat y} = \tfrac{1}{2}\big[(1\pm i)\ket{S{=}1,M_S{=}{-1}} - (1\mp i)\ket{S{=}1,M_S{=}1}\big],\\
  &&\ket{\hat z}= \ket{S{=}1,M_S{=}0}.
\end{eqnarray}
}

\begin{figure*}[h]
\includegraphics[width=.95\textwidth]{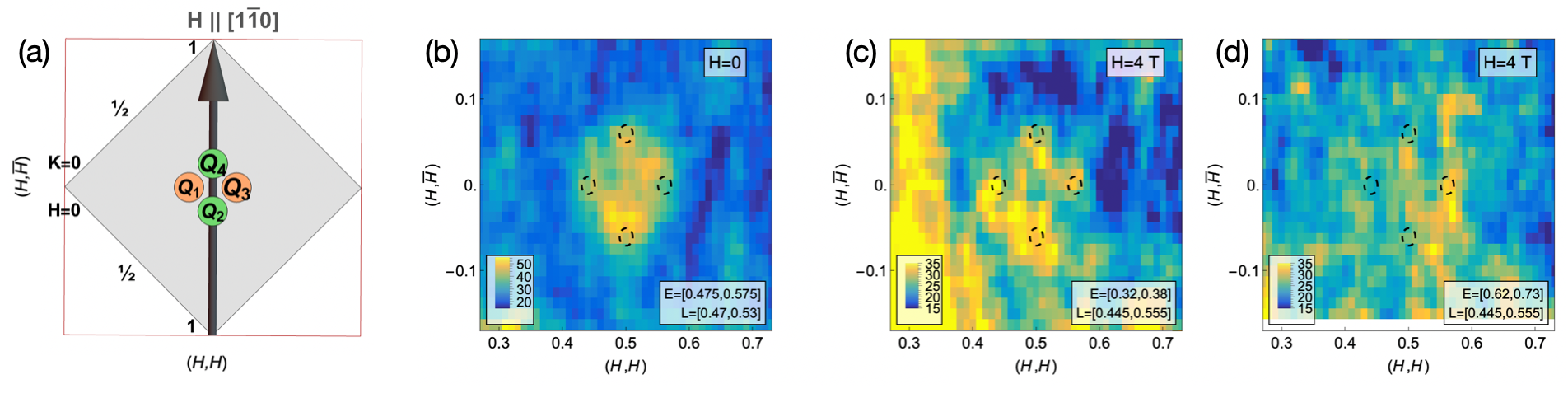}
\caption{ 
\textbf{Wavevectors observed at $H=0$ and $H=4$~T.}
(a) Schematic of the four arms of the star of $Q_1=(0.44,0.44,1/2)$ within the $(H,H,1/2) \times (H,\bar{H},1/2)$ plane.
(b) Zero-field inelastic neutron scattering at IN5 at constant energy $E=0.525 \pm 0.05$~meV.
The dashed circles represent the bottom of the cone-like structures that disperse upward from these positions (see Supplementary information \cite{SI} for details).
(c) Constant-energy slice at $H=4$~T with $E=0.35 \pm 0.03$~meV corresponding to the downward-shifting branch, with all four satellites visible, and (d) $E=0.675 \pm 0.055$~meV.
}
\label{fig2}
\end{figure*}

The resulting amplitudes are evaluated at the four measured modulation
vectors, defined consistently with the experimental scattering geometry (see Fig.~\ref{fig2} (a) and SI \cite{SI} for details). Expressed in reciprocal
lattice units
\begin{align}
\mathbf{Q}_1^{M} &= (q,\, q,\, 0.5), &
\mathbf{Q}_2^{M} &= (1-q,\, q,\, 0.5), \notag\\
\mathbf{Q}_3^{M} &= (1-q,\, 1-q,\, 0.5), &
\mathbf{Q}_4^{M} &= (q,\, 1-q,\, 0.5),
\label{eq:Qvectors}
\end{align}
where $q = 0.44$.

In the high-field regime, once the Q phase develops, $H > H_Q$, the ground state becomes a superposition of the homogeneous spin-singlet $d_{x^2 - y^2}$ state and a PDW carrying spin-triplet correlations. Accordingly, the initial state entering the evaluation of the cross-section evolves into $\ket{\lambda}_{H>H_Q} = \alpha \ket{d_{x^2 - y^2}} + \beta \ket{\lambda_{\rm PDW}}$,
with $\ket{\lambda_{\rm PDW}} = \beta_d |\Psi_d^e\rangle + \beta_{\bar{d}}|\Psi_{\bar{d}}^e\rangle+\beta_z |\Psi_z^e\rangle$, with coefficients $\beta_i$ defined from the eigenvectors of Eq.~\ref{eq:3x3_matrix} in our effective approach. 
The presence of the triplet component
opens a new scattering channel, connecting the PDW modulated component of the ground state to homogeneous spin-singlet excited states. 
Within the $C_{2h}$ point group, descending from $D_{4h}$ under the applied field, the only homogeneous spin-singlet configuration compatible with even parity transforms as the $A_g$ representation, with the simplest case captured by an s-wave order parameter.
Evaluating Eq.~\eqref{eq:INS_matrix} between the PDW component of the ground state and the $A_g$ configuration, we find non-vanishing matrix elements. More details can be found in the SI \cite{SI}.

\section*{Data availability}
The ILL data is available at the cited dois \cite{in5,thales} and will be made available prior to publication.
The ORNL data is available through the Oncat system and will be made available prior to publication.


\bibliography{biblio}
\bibliographystyle{sciencemag}


\section*{Acknowledgments}
A portion of this research used resources at the Spallation Neutron Source, a DOE Office of Science User Facility operated by the Oak Ridge National Laboratory. The beam time was allocated to HYSPEC on proposal number IPTS-28345.1.

\section*{Funding}
This research was supported by the Swiss National Science Foundation (SNSF) under grant 200021\_184983 (M.K.).
D.W.T. acknowledges funding from the European Union’s Horizon 2020 research and innovation programme under the Marie Skłodowska-Curie grant agreement No 884104 (PSI-FELLOW-III-3i). 
I.M. acknowledges financial support by the Swiss National Science Foundation (SNSF) via the SNSF postdoctoral Grant No. TMPFP2\_217204.
A.R. acknowledges financial support from the Swiss National Science Foundation (SNSF) through Ambizione Grant No.PZ00P2\_186043.

\section*{Author contributions}
D.W.T. and M.K. conceived the experimental work and guided the project.
I.M. and A.R. conceived the theoretical description and carried out symmetry analysis and quantitative calculations.
D.J.G. and E.P. prepared samples and made initial characterizations of their physical properties, and D.W.T. coaligned them on aluminum plates in preparation for the neutron work.
S.R., D.W.T., and J.O. conducted the IN5 experiments. 
D.W.T. and B.W. conducted the HYSPEC experiments.
D.W.T., S.R., P.S., and A.H. conducted the Thales experiments.
D.W.T, I.M., A.R., and M.K. prepared the manuscript with input from all authors.

\section*{Competing interests}
The authors declare no competing interests.

\section*{Materials \& correspondence}
Correspondence may be addressed to David W. Tam, Ilaria Maccari, Aline Ramires, or Michel Kenzelmann.



\newpage

\newpage 

\begin{figure*}[h]
\includegraphics[width=.95\textwidth]{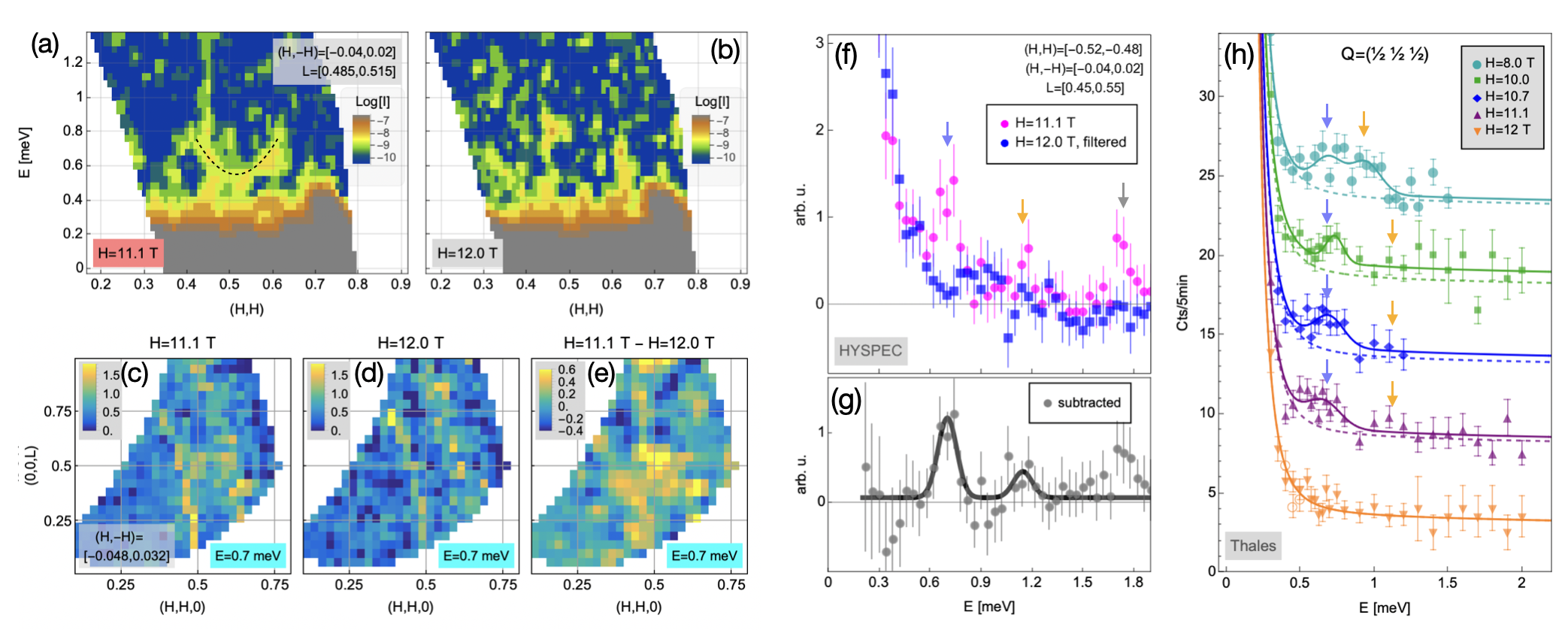}
\caption{ 
\textbf{Inelastic neutron scattering in the high-field regime.}
(a) Inelastic neutron scattering at HYSPEC in the Q-phase at $H=11.1$~T and (b) in the normal state at $H=12.0$~T.
The appearance of an upward-dispersing feature stemming from $Q=(1/2,1/2,1/2)$ is sketched with a dashed line, and is only visible in (a).
(c) Constant-energy slices at $E = 0.7 \pm 0.075$~meV in the $(H,H,0) \times (0,0,L)$ scattering plane at $H=11.1$~T, (d) $H=12.0$~T, and (e) subtracted $H=11.1 - 12.0$~T and smoothed, to enhance the SRM feature appearing at $Q=(1/2,1/2,1/2)$, where the four individual satellites are
overlapping.
(f) Integrated intensity of the SRM from HYSPEC data with a two-point smoothing applied to the $H=12.0$~T data to enhance readability, and (g) subtracted data, showing the presence of a strong feature at $E=0.7$~meV and a weaker feature near $E=1.1$~meV that is consistent with the expected position of the upward-shifting branch of the SRM.
(h) Constant-$Q$ scans at Thales at five fields of $H=8.0$, 10.0, 10.7, 11.1, and 12.0~T.
For $H=8.0$~T our fit assumes the presence of both the central and upper branch, while for $H=10.0-11.1$~T, the upper branch is considerably weaker, and therefore only the central branch is fitted.
The central branch maintains a similar intensity at all fields before disappearing completely at 12.0~T.
The fitted form of the 12.0~T data is therefore assumed to represent the field-independent instrumental background, and is shown as a dashed line under the other data sets.
The purple and orange arrows indicate the positions of the central and upper branches, respectively.
Since the upper branch was not observed well in the Thales experiments, we indicate the expected position while noting that the peak may be absorbed into the presumed background.
}
\label{fig3}
\end{figure*}

\newpage 

\begin{figure*}[h]
\centering
\includegraphics[width=\textwidth]{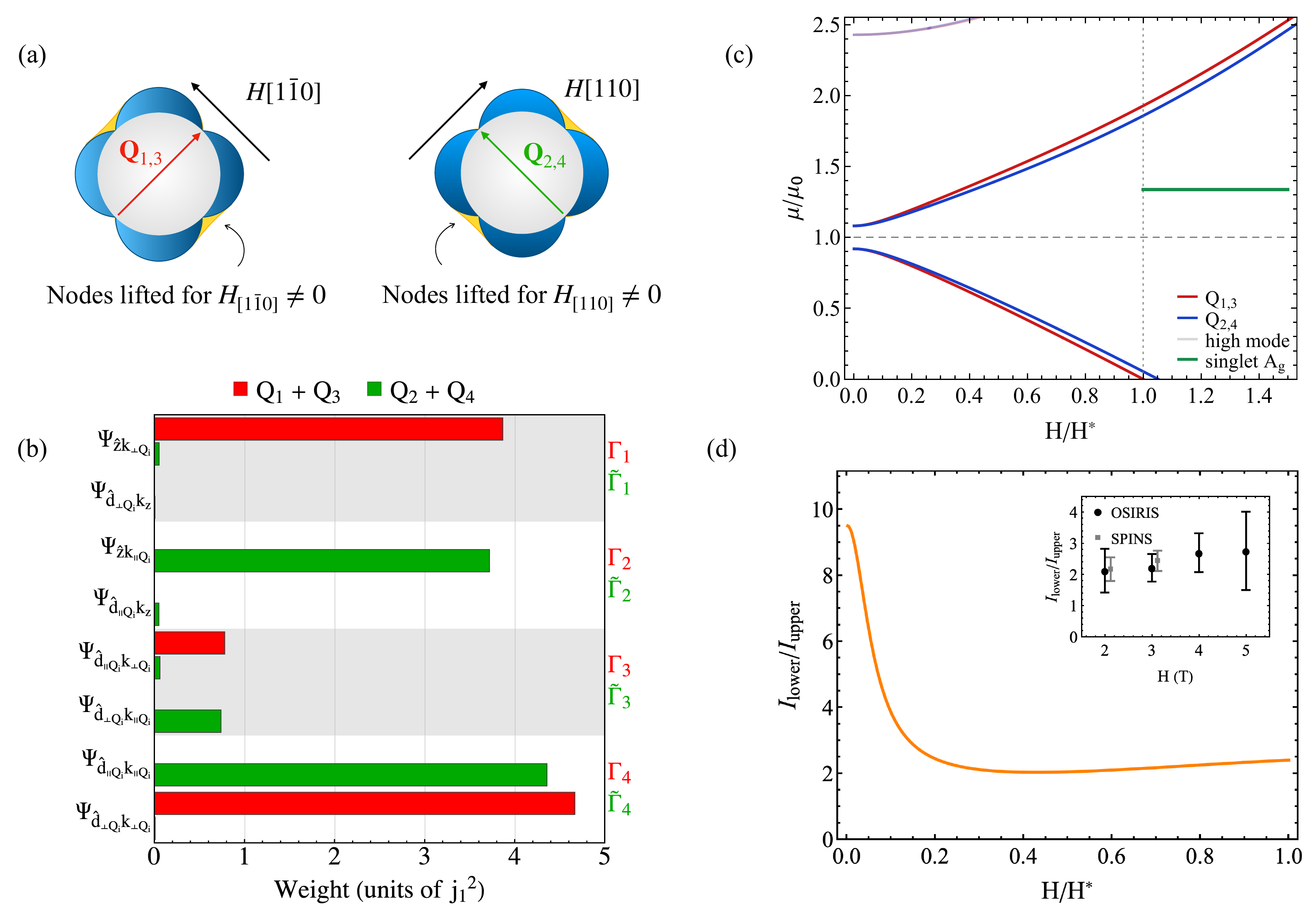}
\caption{ 
\textbf{SRMs as collective excitations of the superconducting state. }
\textbf{(a)} Schematic representation of the nodal structure of the homogeneous superconducting order parameter, $\Psi_{B_{1g}}$, with $d_{x^2-y^2}$ form factor (blue). In the presence of an external magnetic field along the $[1\bar{1}0]$ ($[110]$) direction, the mirror symmetry with normal along the diagonal (antidiagonal) is broken, and the nodes along the magnetic field direction are generally lifted, preempting the development of modulated orders with specific $Q_i$ vectors. 
\textbf{(b)} Neutron-scattering weights of the individual PDW spin-triplet
components for $\mathbf{Q}_1+\mathbf{Q}_3$ (red) and $\mathbf{Q}_2+\mathbf{Q}_4$
(green), grouped by irreducible representation, $\Gamma_i$ for $\mathbf{Q}_{1,3}$
and $\tilde\Gamma_i$ for $\mathbf{Q}_{2,4}$. Form factors are written relative to
each $\mathbf{Q}_i$ (see SI \cite{SI}).
\textbf{(c)} Field evolution of the SRM branch energies from
Eq.~\eqref{eq:3x3_matrix}, at
$\mathbf{Q}_{1,3}$ (red) and $\mathbf{Q}_{2,4}$ (blue), for a representative set of parameters: $\alpha_d=1; \alpha_{\bar{d}}=1.1; \alpha_z= 0.55; \alpha_{d\bar{d}}=0.4;\lambda=0.7; \gamma=0.3$. Energies are in units of
$\mu_0$, the zero-field center of the split pair of modes (see SI for more details), and $H$ in units of
$H^*$, where the lower branch softens to zero.
$H$ in units of $H^*$, where the lower branch softens. 
\textbf{(d)} Intensity ratio of the lower to the upper branch for the same
parameters as in (c). Inset: experimental data point from Ref.~\cite{Stock2012}.
(See Supplementary information \cite{SI} for details).
}
\label{fig4}
\end{figure*}

\end{document}